\documentclass{article}

\usepackage{arxiv}

\usepackage[utf8]{inputenc} 
\usepackage[T1]{fontenc}    
\usepackage{hyperref}       
\usepackage{url}            
\usepackage{booktabs}       
\usepackage{amsfonts}       
\usepackage{nicefrac}       
\usepackage{microtype}      
\usepackage{lipsum}
\usepackage{graphicx}
\graphicspath{ {./images/} }

\title{The Security of Quantum Computing in 6G: from Technical Perspectives to Ethical Implications}

\author{
 Luca Barbieri\\
  University of Genoa\\
  Genoa, Italy \\
  \texttt{barbieriluca.bl@gmail.com} \\
   \And
 Abdelkrim Menina\\
  Deutsche Telekom Chair of Communication Networks\\
  Technische Universität Dresden\\
  Dresden, Germany \\
  \texttt{abdelkrim.menina@tu-dresden.de} \\
  \And
 Riccardo Bassoli\\
  Deutsche Telekom Chair of Communication Networks\\
  Technische Universität Dresden\\
  Dresden, Germany \\
  \texttt{riccardo.bassoli@tu-dresden.de} \\
  \And
Frank H.P. Fitzek\\
  Deutsche Telekom Chair of Communication Networks\\
  Technische Universität Dresden\\
  Dresden, Germany \\
  \texttt{frank.fitzek@tu-dresden.de} \\
}

\begin{document}
\maketitle
\begin{abstract}
Quantum technologies hold promise as essential components for the upcoming deployment of the future 6G network. In this future network, the security and trustworthiness requirements are not considered fulfilled with the current state of the quantum computers, as the malicious behaviour on the part of the service provider towards the user may still be present. Therefore, this article provides an initial interdisciplinary work of regulations and solutions in the scope of trustworthy quantum computing for future 6G that can be viewed as complimentary regulations to the existing strategies shared by different actors of states and organizations. More precisely, we describe the importance of a reliable quantum service provider and its implication on the ethical aspects concerning digital sovereignty. By exploring the critical relationship between trustworthiness and digital sovereignty in the context of future 6G networks, we analyse a trade-off between accessibility to this new technology and preservation of digital sovereignty engaging in parallel the United Nation's (UN's) sustainable development goals. Furthermore, we propose a partnership model based on cooperation, coordination, and collaboration giving rise to a trusted, ethical, and inclusive quantum ecosystem, whose implications can spill over to the entire global scenario.
\end{abstract}

\setcounter{tocdepth}{2}
{\small
\tableofcontents}

\section{Introduction}

\begin{figure}[h!]
    \centering
    \includegraphics[width=0.9\linewidth]{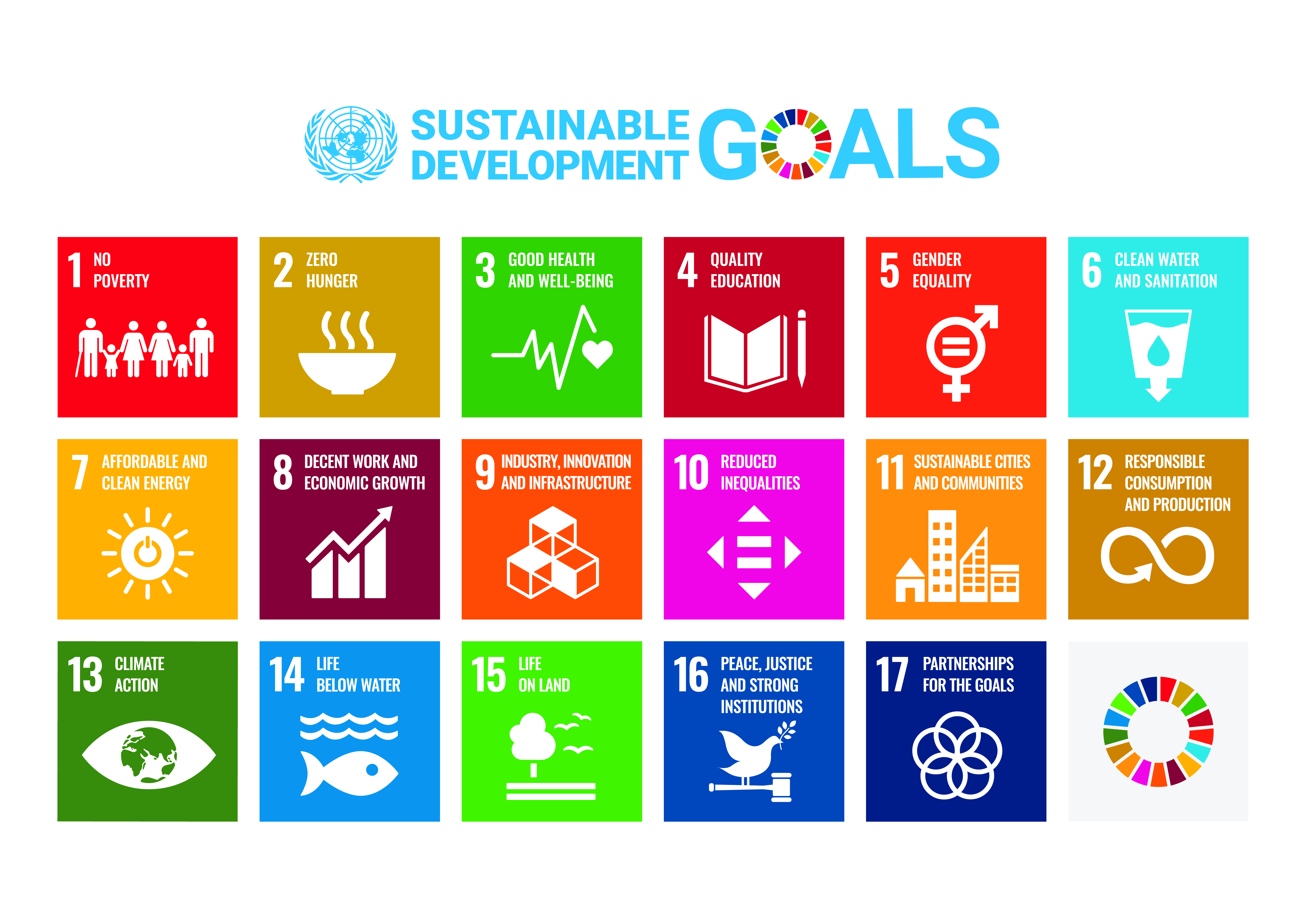}
    \caption{The United Nation's seventeen Sustainable Development Goals \cite{SDGs-17}.}
    \label{SDGs_17}
\end{figure}
Recent advances in quantum technologies are bringing us closer to a profound shift in science and technology that will have far-reaching implications for our economies, security, and defence.
These technologies could revolutionise sensing, imaging, precise positioning, navigation and timing, communications, computing, modelling, simulation and information science \cite{Quantum_Sensing, QKD_Rubenok_2013, VBQC_Kashefi}.
The advent of the quantum computer has brought with it great innovations in the world of information, enabling operations that a classical computer cannot perform \cite{manin2000classical, QML, cordier2022biology}. This significant growth of quantum technologies has caught the attention of the telecommunication community, especially within the context of emerging 6G networks \cite{Quantum-Enabled_6G_Wang, Survey_Beyond_5G, bassoli2021quantum, book_Frank_2020}, attention further stimulated by the quantum internet concept with its premises on the quantum internet \cite{Quantum_internet_Stephanie}.

The arrival of the future generation of 6G networks in 2030 coincides with the agenda of achieving the United Nation's sustainable development goals (SDGs). The goals include 17 objectives (as depicted in figure \ref{SDGs_17}) in a broad program of action for a total of 169 targets or objectives. The official launch of the SDGs corresponded with the beginning of 2016, guiding the world on the path to follow in the next 15 years; the countries, in fact, have committed to achieving them by 2030. The development goals follow the results of the millennium development goals that preceded them and represent common objectives on a series of important development issues: fighting poverty, eliminating hunger, and fighting climate change, to name just a few. “Common goals” means that they concern all countries and all individuals; no one is excluded, nor should they be left behind in the journey necessary to bring the world on the path to sustainability. 

With the current state of quantum computers, the 6G requirement of trustworthiness is not seen as satisfied, making the achievement of the SDGs unmet. Indeed, the service provider may exhibit malicious behaviour compromising. A vulnerability arising from a lack of digital sovereignty.

Therefore, in this work we highlight a potential trade-off between the SDGs and tackle this issue by providing technological and political regulations. The paper is structured as follows:
\begin{enumerate}
    \item First, in Section \ref{Strategies} we recall the current strategies of different actors regarding the digital sovereignty of quantum technologies.
    \item Then, Section \ref{Ethical} explores the existing connection between trustworthiness and digital sovereignty in the future 6G, while emphasizing its implication on the ethical level, providing an example of a trade-off between SDGs when the ethical responsibility of the service provider is absent.
    \item Finally, in Section \ref{Regulations} we analyze new prospectives of regulations on quantum computing to overcome this trade-off based on technological and political solutions, followed by a discussion in Section \ref{Discussion} to discuss the extension of the 3C model composed of cooperation, coordination, and collaboration to quantum computing in future 6G.
\end{enumerate}

\section{Existing Perspectives on Providing Strategies for Digital Sovereignty \label{Strategies}}
A sovereign state is an independent state, ``recognized within its borders by the international community" and which exercises over its people ``a power of administration and jurisdiction" following the definition in Larousse \cite{Sovereignty_Larousse}. Referring to the digital sphere, this notion is much more difficult to circumscribe. If Digital Sovereignty globally indicates the fact, for a state or an organization, of asserting its authority in order to exercise its prerogatives within cyberspace, it also responds to more concrete problems, such as technological dependence or control of the user's personal data.\\
According to the World Economic Forum \cite{WEF}, Digital Sovereignty refers to the ability to have control over your digital destiny, that is, the data, hardware, and software you rely on and create. This ability is crucial to ensuring data security and, in turn, the protection of citizens and businesses from potential cyber threats.
Therefore, it has become a concern for many policy makers who believe that there is too much control ceded to too few places, too little choice in the technology market, and too much power in the hands of a small number of large technology companies.

To achieve Digital Sovereignty it is essential that each state shares a Regulatory Strategy on Quantum Technologies. At present, not many countries have enacted a strategy, as it is a developing technology for which it is difficult for legislators to have a long-term vision. However, it is possible to mention a few countries that have issued a Quantum Technologies Regulatory Strategy. Specifically, in Ireland, the Department of Further and Higher Education, Research, Innovation and Science published in November 2023 a complete strategy document \cite{Quantum_strategy_Ireland} with the aim of becoming an internationally competitive hub in the field of quantum technologies, at the forefront of scientific and engineering advances, through research, talent, collaboration, and innovation.\\
This vision is based on five pillars: Pillar 1 focuses on supporting excellent fundamental and applied quantum research. This internationally excellent research supports advances in quantum technologies. Ireland has a strong track record in fundamental and applied quantum research and aims to further improve it to enable breakthrough discoveries and feed the pipeline of innovations and technologies. Pillar 2 highlights that the best way to circulate knowledge in the economy and society is to foster the best talent in science and engineering. Meanwhile, pillar 3 aims at increasing national and international collaborations, often involving partners from multiple sectors. Ireland already participates in several European academic collaborations on quantum technologies and related areas. To fully exploit the opportunities that quantum technologies will offer, the Ireland strategy emphasises in pillar 4 the importance of focusing on innovation, entrepreneurship, and economic competitiveness. This pillar seeks to stimulate innovation and entrepreneurship in quantum technologies and related areas, including indigenous small and medium-sized enterprises. It also aims to strengthen collaborative work between academia and business. Finally, as quantum technologies are new and rapidly evolving, it is important to spread awareness of quantum technologies and their concrete benefits among a wide range of stakeholders, as stated in pillar 5. The aim of this pillar is to have a quantum-literate society that takes full advantage of the benefits that quantum technologies can bring for all. 
To ensure that the ambitions of pillars 1 to 4 are realised, stakeholders involved in these pillars will also need to focus on the actions outlined in pillar 5.

The United Kingdom's Department for Science, Innovation $\&$ Technology, in March 2023, shared its national quantum strategy, which sets out a vision and 10-year plan for quantum technologies in the UK, pledging to spend £2.5 billion on research, innovation, skills, and more \cite{Quantum_strategy_UK}. The UK's vision is to become a leading quantum technology-based economy by 2033, with a world-leading sector in which quantum technologies are an integral part of the UK's future digital infrastructure and advanced manufacturing base, stimulating growth and helping to build a thriving and resilient economy and society. This long-term commitment builds on the successful foundations laid by the National Quantum Technologies Program. In line with the UK national strategy, this program will continue to invest in the facilities and infrastructure in which to build, host, and operate quantum computers, while supporting the training and development of the talent and skills pipeline and growing the industrial ecosystem nationally.

The Canadian National Quantum Strategy \cite{Quantum_strategy_Canada} defines three key missions to ensure that Canada remains on the path of quantum innovation and leadership: first, to make Canada a world leader in the development, deployment and continued use of quantum computing hardware and software for the benefit of Canadian industry, governments and citizens. Secondly, to ensure the privacy and cybersecurity of Canadians in a quantum-enabled world through a secure national quantum communications network and a post-quantum encryption initiative. Finally, enabling the Canadian government and key industries to be developers and early adopters of new quantum sensing technologies.

In Australia, the National Quantum Strategy has 5 central themes \cite{Quantum_strategy_Australia}. Each theme has a set of actions over 7 years that will position Australia for success and are focused on: (i) creating thriving research and development by investing in quantum technologies; (ii) securing access to essential quantum infrastructure and materials; (iii) building a skilled and growing quantum workforce; (iv) ensuring our standards and frameworks support national interests; (v) building a trusted, ethical and inclusive quantum ecosystem.
In addition, the strategy outlines how the Australian government will deliver on its vision. It also signals areas the government may consider in the future, including investment opportunities.\\
Australia’s quantum ambition will not be achieved by working alone; instead, every part of the quantum ecosystem needs to work towards the same goal, including through investments. The government will drive the implementation of the strategy, but other partners will lead some actions and initiatives.

The US government in 2022 issued a presidential memorandum \cite{Quantum_strategy_Biden} setting out the strategy and a list of guidelines to be followed by the various agencies in the transition to the new quantum algorithms. The term ‘agency’ in this case means every executive department, military department, government corporation, government-controlled corporation and every operational branch of government.\\
By 4 May 2023, one year after the President's memo, all agencies had to send the OMB an inventory of systems and assets containing cryptographic systems vulnerable to quantum attacks. In May 2024, a second inventory was also compiled regarding cost estimates for the transition of affected cryptographic systems.\\
As a conclusion and validation of this process, the US government passed Public Law 117-260 on 21 December 2022 containing the ‘Quantum Computing Cybersecurity Preparedness Act’ \cite{Quantum_strategy_U.S_gov}, in which the previous notes are reiterated and put into effect, reaffirming the importance and urgency of activating the migration procedures to post-quantum algorithms, with the goal of completing it by 2035.

In January 2024, NATO published its first quantum technology strategy to establish a quantum-ready alliance\cite{Quantum_strategy_NATO}. More precisely, NATO foreign ministers approved the strategy on November 28, which highlights quantum technologies as a key element of strategic competition, presenting unprecedented advantages in computing, communications, and detection. In detail, the strategy document highlights the dual-use nature of quantum technologies and aims to rapidly and responsibly increase the Alliance's quantum capabilities to prevent adversaries and competitors from gaining a strategic advantage. Although much of the decision-making community has focused on exploiting the capabilities of artificial intelligence and machine learning (AI/ML) for national security while simultaneously preventing its malicious use, the same level of attention should be provided to accelerate technical progress in quantum computing making the strategy both timely and necessary.\\
To become an Alliance ready for quantum technology, NATO and the Allies are promoting the strategy of development of a secure, resilient and competitive quantum ecosystem capable of responding to the fast pace of technological competition in the quantum field. This requires coherence in investments, cooperation between Allies in technology development opportunities, development and protection of skilled workforce and increased situational awareness and information sharing. In addition, NATO's strategy is providing the main transatlantic forum for quantum technologies in defence and security, helping to continuously develop our shared understanding and harness the potential of quantum technologies while protecting them from adversary use.\\
Within this strategy, a ‘learn by doing’ approach is adopted to integrate quantum technology considerations into the implementation of our operational concepts, defence planning cycles, capability development cycles and standardisation efforts. As DIANA \cite{Quantum_strategy_Diana} and the NATO Innovation Fund (NIF) become fully operational, their deep-tech activities will also influence NATO's strategic approach to quantum technologies and strengthen NATO's engagement in the allied quantum ecosystem.

On the Asian continent, several countries have started taking early measures to anticipate quantum threats. China is one of the most strongly active nations in the field of quantum cryptography. In particular, in 2011, the National Space Science Center of the Chinese Academy of Science initiated the ``Strategic Priority Programme on Space Science" \cite{cryptoeprint:2019/510} whose main goals include carrying out experiments on a global scale concerning the quantum phenomena of entanglement and quantum transport and developing a wide-area quantum communication network. Meanwhile, in 2023 Russia collaborated with the Chinese quantum communication network, carrying out quantum communication between two earth stations 3800 kilometres apart \cite{Quantum_strategy_China-Russia}.

Even though different countries provide different strategies for future deployment of quantum technologies, only Australia mentioned the central theme of building a trusted, ethical, and inclusive quantum ecosystem. To the best of our knowledge, the security of quantum computing in 6G is not specified in the afforementioned strategies. Therefore, we propose an initial framework to lay the groundwork on digital sovereignty based on interdisciplinary collaboration.

\section{Ethical Responsibility of the Quantum Computing Service Provider Towards the Consumer\label{Ethical}}

Ethical implications of responsible research and innovation have been envisioned in \cite{kop2023towards}, which provides an initial framework for a responsible quantum ecosystem and guidelines for researchers and industrials to address future challenges concerning economic security, dual use, privacy, product safety and liability, fair competition, and equality. Challenges that are highly related to digital sovereignty, as discussed before.\\
Achieving these challenges means sharing the same objective as the UN's SDGs. Therefore, in this section, we first describe the relationship between trustworthiness and digital sovereignty regarding quantum computing. Then we outline the contribution of quantum computing to the SDGs and highlight an example of a trade-off between the SDGs. 

\subsection{Trustworthiness and Digital Sovereignty as a Requirement for Future Deployment of Quantum Computing in the 6G \label{Trustworthiness}}

According to the fact that big data is correlated with the unique patterns of individual behaviour, removing related personal information might not safeguard the individual's privacy. Persons or groups of persons can be re-identified when databases are reconstituted. In more general terms, such divulgations are a result of weak data and digital sovereignty; while the former is about control, the latter concerns ownership.

In the domain of quantum computing, many companies are providing cloud solutions, and the safety of the consumer's data is as compromised as for classical cloud computing. The reason behind the cloud solution is the expensiveness of affording quantum computers, since this technology is still at its early age of development. In this regard, the individuals have no lead over the governance of the infrastructure and neither on the collection nor storage of private data, thereby generating a trustworthiness issue. 

In \cite{ylianttila20206gwhitepaperresearch}, trust in a network context is described as the ``expected outcomes of communicating with a remote party in a session. The possible outcomes are either the positive value of the communication or being hacked or cheated in some way.". In other words, trustworthiness is defined as the assurance that a system will perform as expected, and without this requirement, scientific and technological progress can have unintended negative consequences.\\
At the current stage, quantum computers are far from reaching their applicative potential as envisaged by researchers since they are subject to errors and instabilities. In such a stage, one might cease this opportunity to develop a common responsibility for a desirable ethical and societal outcome.

\subsection{Impact of Quantum Computing on Sustainable Development Goals and Trade-off}

It is important that quantum computing can help us achieve the 17 United Nations SDGs more quickly, creating a more secure and equitable world for current and future generations. Although this technology was not mentioned in 2015, quantum computing technology may be the accelerator to make giant steps before the 2030 deadline \cite{WEF}. Some domains to which quantum computing is expected to contribute with significant environmental and societal opportunities vary from molecular simulation and discovery in materials science and biology to optimization and risk management in complex systems. In the latter contribution, for example, Ford Motors investigated quantum-inspired strategies to alleviate traffic congestion in Seattle as part of a recent drive to reduce environmental impacts in metropolitan areas and promote sustainable cities (aligned with SDG 11) \cite{Ford}. When compared to conventional'selfish' routing tactics, the trial's optimal routing recommendations resulted in a $73\%$ decrease in total congestion.

\begin{figure}[!t]
    \centering
    \includegraphics[width=0.5\linewidth]{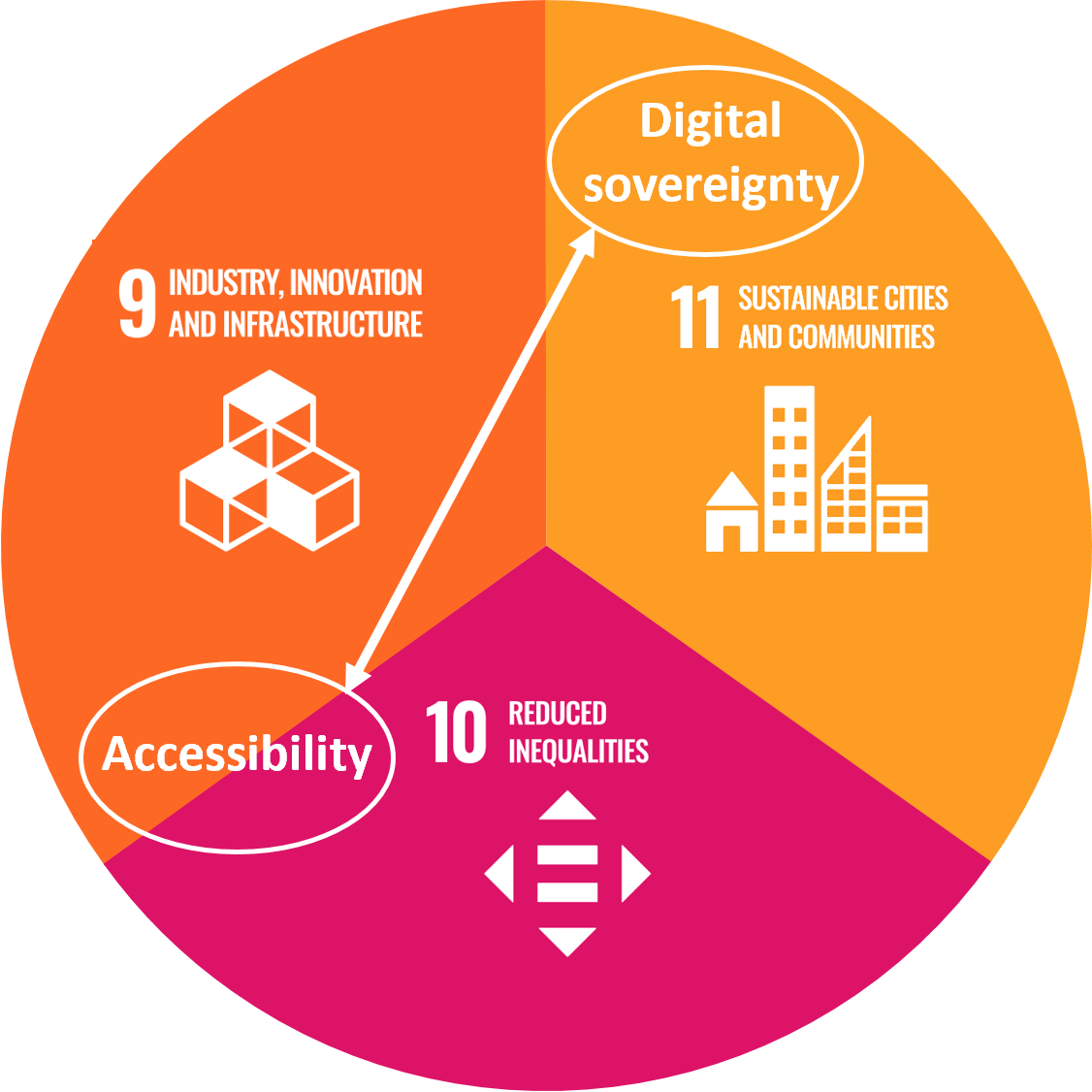}
    \caption{First trade-off between SGDs.}
    \label{Trade off}
\end{figure}

Previous work has highlighted SDGs trade-off between eradicating hunger and malnutrition in goal 2 and the protection of the biodiversity ecosystem in goal 15. As food production increases, it generates greenhouse gas emissions from harnessing water sources, land, and energy \cite{mbow2020food}. Under our framework, in figure \ref{Trade off} is presented the trade-off between different SDGs. Digital sovereignty is required to achieve goal 11 of \textit{Sustainable Cities and Communities}, in the sense that it is expensive to acquire a quantum computer, but the overall benefit is to insure digital sovereignty. As the quantum computers are already available via cloud, users from different sectors have remote accessibility to their personal usage, therefore contributing to goals 9 and 10. In other words, the cloud provides accessibility for decreasing the inequalities; on the other hand, this comes with the expense of losing digital sovereignty.

However, in order not to make the mistake of regulating ex post a technology that has been implemented, it is clear that there is a need for many countries to regulate quantum technology in order to achieve digital sovereignty.

\section{Proposed Solutions and Regulations\label{Regulations}}

\begin{figure}[!t]
    \centering
    \includegraphics[width=0.7\linewidth]{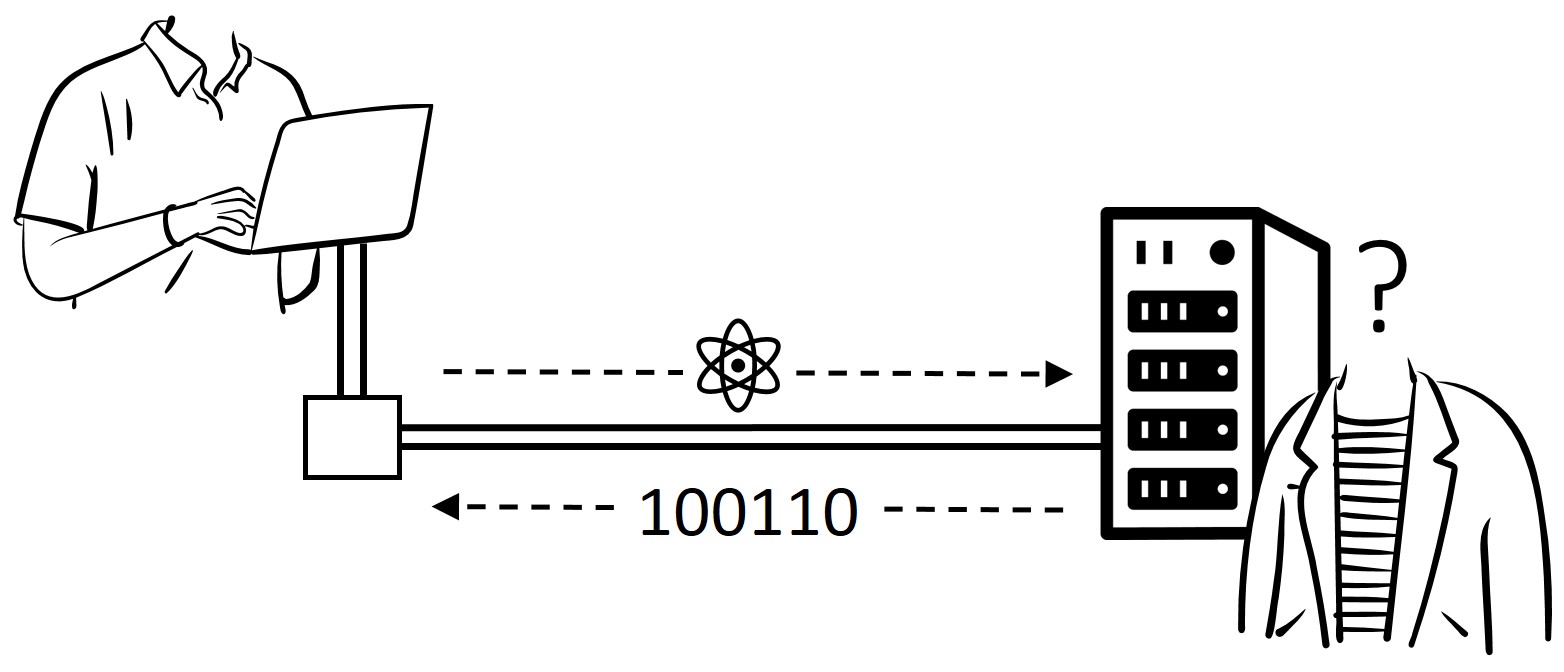}
    \caption{Basic scheme of blind quantum computing. On the left side the client and on the right side the malicious server.}
    \label{BQC}
\end{figure}

To protect the user while at the same time making quantum computing accessible, we propose some existing technical solutions at the European level to regulate cloud quantum computing. Furthermore, we propose to extend the existing bill of the Italian Senator Lorenzo Basso to include another measure \cite{Senatore_Basso}. 

Such technical solutions have been thought of in the community of Quantum Internet Alliance (QIA), working together to build the quantum internet network based on quantum blind computing. This method of quantum computing is based on blinding the server and following the protocol as described in figure \ref{BQC} 
\begin{enumerate}
    \item Preparation of quantum states (client side) and transmission of qubits to the server.
    \item Server constructs a quantum circuit.
    \item Measurements and execution of Computation (server side).
    \item Feedback to the client with classical information then decoding.
\end{enumerate}
Beside QIA, the Schirmprojekt Quantenkommunikation Deutschland (SQuaD) \cite{SQUAD} plays a key role in ensuring that basic research and industry are optimally networked and connected in the coming years. Its goal is transferring current Quantum Communication Technologies sustainably into industrial applications. From SQuaD standardization workshops, potential solutions for identified needs were developed regarding the protection profile, highlighting missing standards for hardware authentication/validation and its implementation. Under the umbrella of authentication, quantum physical unclonable functions (QuPUFS) are presented as potential candidates \cite{phalak2021quantum}.

These solutions facilitate the technological transition by improving digital sovereignty and stimulating economic growth through independent and secure technological innovations. Commitments missing in the bill that Italy submitted in April 2024 to allocate up to EUR 1 billion to companies operating in the fields of AI, cybersecurity, quantum computing and advanced telecommunications.\\
The existing measures envisaged by the bill are divided into six macro areas: 1) Development and Research: Investment of €1 billion over five years to strengthen the National Research Centre in High Performance Computing, Big Data and Quantum Computing, expanding infrastructure and promoting participation in grants and the funding of doctoral scholarships; 2) National Coordination: Establishment of the National Quantum Coordination to facilitate collaboration between government, universities, research centres and companies, and to promote access to quantum technologies; 3) Internationalisation: Creation of a €20 million annual fund to foster international collaborations, strengthening global networks and taking a leading role in multilateral fora; 4) Start-ups and Innovation: Launch of the ‘Quantum Acceleration Fund’ with an endowment of €300 million over five years to support the development of innovative start-ups in the field of quantum technologies, accelerating the commercialisation and industrialisation of technological solutions; 5) National security: Overhaul of strategic infrastructures to make them ‘quantum-safe’ with an investment of EUR 50 million, ensuring greater protection against cyber threats; 6) Training and Skills: Allocation of €100 million to train a new generation of experts in quantum technologies, promoting multidisciplinary programmes and research opportunities at university and post-doctoral level. As an extension, a seventh measure would concern the protection of the user from a susceptible malicious quantum computing service provider through ensuring accessibility to the technological device.

\section{Discussion\label{Discussion}}

The technical solutions and regulations are relevant only if different partners at the global level accomplish cooperation, collaboration, and coordination as stated in \cite{fratini2024digital}. The virtuous governance model on which cloud quantum computing in 6G should be based on the 3C model of i) Cooperation: cloud quantum computing in future 6G can be the means to structure cooperation between allied countries aimed at strengthening digital sovereignty; ii) Collaboration: strengthening global networks with public-private partnerships is the right method for each country; iii) Coordination: making efficient strategic through organized infrastructures, especially critical ones, to achieve digital sovereignty is a priority for which international coordination is required. By pursuing these three objectives as a model will it be possible to give rise to a trusted, ethical and inclusive quantum ecosystem.

Also in relation to the global scenario strategy, it is noteworthy to mention how in September 2024, the U.S. Department of Commerce’s National Institute of Standards and Technology (NIST) has finalized its principal set of encryption algorithms designed to withstand cyberattacks from a quantum computer \cite{Quantum_strategy_NIST}. Researchers around the world are racing to build quantum computers that would operate in radically different ways from ordinary computers and could break the current encryption that provides security and privacy for just about everything we do online. To defend against this vulnerability, the recently announced algorithms are specified in the first completed standards from NIST’s post-quantum cryptography (PQC) standardization project and are ready for immediate use. Steps are taken towards protecting the quantum service provider against such attacks, which implies a thorough study of its implications for the SDGs.

\section{Conclusion}

Insuring a trustworthy service provider for quantum computing is essential for the future 6G communication networks. We have demonstrated that this technology can benefit society only if the service provider is ethically responsible and establishes a trustworthy connection with the user. In the worst-case scenario, where the service provider fails to meet this requirement, a trade-off arises between digital sovereignty and the accessibility of these new technologies, thereby impacting compliance with the Sustainable Development Goals. Therefore, technological regulations based on blind quantum computing and QuPUFs are necessary to resolve this trade-off, but a measure needs to be taken at the political level, as shown with the example of the bill proposed by the Italian policymaker.\\
Finally, we arrive at a modified 3C model applied to quantum computing in 6G. Adherence to this model, based on cooperation, collaboration, and coordination, gives rise to a trusted, ethical, and inclusive quantum ecosystem, whose implications can spill over to the entire global scenario.

\section{Acknowledgments}
This project was funded by the German Research Foundation (DFG, Deutsche Forschungsgemeinschaft) as part of Germany’s Excellence Strategy – EXC 2050/1 – Project ID 390696704 – Cluster of Excellence “Centre for Tactile Internet with Human-in-the-Loop” (CeTI) of Technische Universität Dresden. The authors acknowledge also the financial support by the Federal Ministry of Education and Research of Germany in the programme of “Souverän. Digital. Vernetzt.”. Joint project 6G-life, project identification number: 16KISK001K. And in the project Q-TREX, project identification number: 16KISR027.

\label{sect:acks}

\bibliographystyle{unsrt}
\bibliography{references}


\end{document}